\documentclass{article}
\usepackage{spconf}
\usepackage{amsmath,amssymb,amsfonts}
\usepackage{graphicx}
\usepackage{textcomp}
\usepackage{xcolor}

\usepackage{algorithmic}
\usepackage{algorithm}
\usepackage{hyperref}

\usepackage{pgfplots}

\DeclareMathOperator*{\argmax}{argmax}
\DeclareMathOperator*{\argmin}{argmin}

\newcommand{\new}[1]{{\leavevmode{#1}}}

\begin{document}

%TODO: IEEE Transaction is not a conference title

\title{Privacy-Aware Design of Distributed MIMO ISAC Systems \vspace{-0.3cm}}

% \author{\IEEEauthorblockN{Henrik Åkesson, Diana P. M. Osorio, Marco Gomes}
% \IEEEauthorblockA{\textit{Communication Systems}, \textit{Linköping University}, Linköping, Sweden \\
% Emails: {\{henrik.akesson, diana.moya.osorio\}@liu.se}, Marco-gomes.email@portugal.pt}
% }

\newcommand{\resizeaddress}[1]{%
  {\footnotesize
  #1
  }%
}

\name{Henrik Åkesson\textsuperscript{*}, Marco Gomes\textsuperscript{\textdagger}, Diana P. M. Osorio\textsuperscript{*} \vspace{-0.3cm}}
\address{\normalsize{\textsuperscript{*}Divison of Communication Systems, Department of Electrical Engineering, Linköping University, Linköping, Sweden} \\
         \normalsize{\textsuperscript{\textdagger}Instituto de Telecomunicações, Department of Electrical and Computer Engineering, University of Coimbra, Coimbra, Portugal} \\
\normalsize{E-mail: \{henrik.akesson, diana.moya.osorio\}@liu.se, marco@co.it.pt}}

% \oneauthor
%  {Henrik Åkesson\*, Marco Gomes, Diana P. M. Osorio\*}
% 	{\small\shortstack{Linköping University, Linköping, Sweden\\
%     Department of Electrical Engineering \\
% 	Communication Systems\\
% 	Email: \{henrik.akesson, diana.moya.osorio\}@liu.se, marco@co.it.pt}}
% {Marco Gomes}
% 	{\small\shortstack{University of Coimbra, Coimbra, Portugal\\
%     Department of Electrical and Computer Engineering \\
% 	Instituto de Telecomunicações \\
% 	marco@co.it.pt}}

\maketitle

\begin{abstract}
Integrated Sensing and Communication (ISAC) systems raise unprecedented challenges regarding security and privacy since related applications involve the gathering of sensitive, identifiable information about people and the environment, which can lead to privacy leakage. Privacy-aware measures can steer the design of ISAC systems to prevent privacy violations. Thus, we explore this perspective for the design of distributed massive multiple-input multiple-output ISAC systems. For this purpose, we introduce an adversarial model where a malicious user exploits the interference from ISAC signals to extract sensing information. To mitigate this threat, we propose an iterative privacy-aware framework of two blocks: precoder design and access point selection. The precoder design aims to minimize the mutual information between the sensing and communication signals by imposing constraints on sensing and communication performance and maximum transmit power. The access point selection also aims to minimize the mutual information between communication and sensing signals by strategically selecting access points that transmit ISAC signals, and sensing receivers. Results show a reduction in the effectiveness of the attack measured by the probability of detection of the attacker.
\end{abstract}
%Results show a privacy gain is possible when the framework is applied to a simulated network.

\section{Introduction}
The evolution of multiple-input multiple-output (MIMO) technologies toward distributed implementations, will serve as support infrastructure for distributed integrated sensing and communication (ISAC) systems and networked sensing~\cite{11021487}. While significant progress has been done to propose the best strategies for conciliating communication and sensing requirements~\cite{behdad-multi-static-2023}, a primary concern in the design of the physical layer of ISAC is security and privacy \new{\cite{11017670}}. Particularly, the ability of malicious parties to use sensing information to track and locate targets can lead to the emerging of unprecedented attacks \new{\cite{wang_anti-malicious_2024}}. Several methods have been proposed to address the security concerns inherent in ISAC systems. For instance, eavesdropping attacks have been addressed through a number of physical layer security-based safeguards~\cite{9737364,10587082}. For sensing based on radio-frequency signals, protection against human activity recognition from channel state information (CSI) has been widely addressed~\cite{9684968}. However, privacy attacks at the physical layer, along with corresponding countermeasures remain widely unexplored.  In~\cite{da-silva-multi-static-2023}, a privacy attack model is introduced in which an internal adversary exploits the signals received from ISAC access points (APs) to infer the directions of the APs’ sensing beams and thereby estimate the location of sensing targets.

% A privacy attack model is proposed in~\cite{da-silva-multi-static-2023}, which consider an internal attacker that leverages the signals received from ISAC access points (APs) to estimate the position of sensing targets by inferring the directions the APs' sensing beams.

Herein, we propose a modification of the model presented in\cite{da-silva-multi-static-2023} to more accurately consider the distributed MIMO scenario. Furthermore, we propose an iterative framework of two blocks, precoder design and AP selection, to reduce the impact of the attack. The aim of the framework is to minimize the mutual information (MI) between the users' communication signal and the sensing signal. In particular, the AP selection stage determines whether an AP should transmit or receive ISAC signals, thereby strategically controlling information leakage.

\vspace{-0.5cm}

\section{System Model}
\begin{figure}[bt]
    \centering
    \includegraphics[scale=0.25]{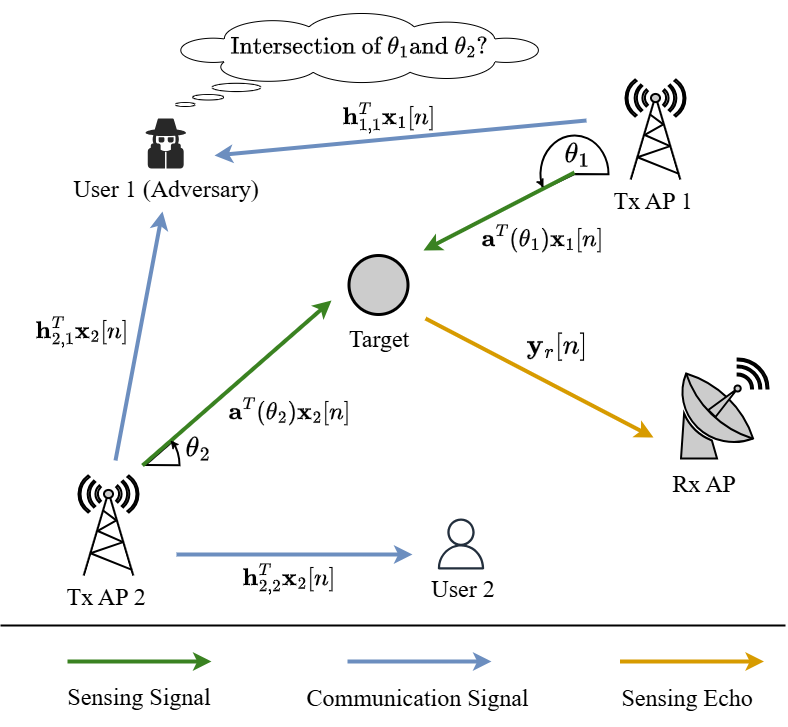}
    \caption{System model of an adversary trying to locate the target by finding the intersection of lines with different angles.}
    \label{fig:system-model}
\end{figure}

Consider a centralized distributed massive MIMO system of $N_{\text{Tx}}$ transmitting APs having $M$ antennas each, jointly serving $N_{\text{UE}}$ single-antenna users. Additionally, $N_{\text{Rx}}$ sensing receiver APs of a static target present in the system, which is illuminated by the ISAC transmitters. In addition, this system model also includes an adversarial user trying to locate the target by estimating the sensing beam angles as illustrated in Fig.~\ref{fig:system-model}. With $n $$=$$ 1,2...,N$ denoting the $n$th time sample, the received signal, $y_i[n]$, at user $i=1,2,...,N_{\text{UE}}$ with communication symbols $s_i[n]$, is given by
\vspace{-0.3cm}
\begin{multline}\label{eq:received-signal}
    y_i[n] = \underbrace{\sum\limits_{j=1}^{N_{\text{Tx}}}\mathbf{h}_{j,i}^H\mathbf{w}_{j,i}s_i[n]}_{\text{Desired Signal}} + \underbrace{\sum\limits_{j=1}^{N_{\text{Tx}}}\sum\limits_{\substack{k=1 \\ k\ne i}}^{N_{\text{UE}}}\mathbf{h}_{j,i}^H\mathbf{w}_{j,k}s_k[n]}_{\text{Communication Interference}} \\ 
    + \underbrace{\sum\limits_{j=1}^{N_{\text{Tx}}}\mathbf{h}_{j,i}^H\mathbf{w}_{j,t}s_s[n]}_{\text{Sensing Interference}} + \underbrace{n_i[n]}_{\text{Noise}},
\end{multline}

\noindent where $\mathbf{h}_{j,i}$ and $\mathbf{w}_{j,i}$ are $M\times1$ column vectors for the channel coefficients and the precoder, from AP $j$ to  user $i$, respectively. Similarly, $\mathbf{w}_{j,t}$ is the sensing precoding vector from AP $j$ to the target $t$. The channel vectors $\mathbf{h}_{j,i}\in\mathbb{C}^{M}$ are independent and identically distributed (i.i.d.) $\mathcal{CN}(\mathbf{0},\beta_{j,i}\mathbf{I})$, where $\beta_{j,i}$ is the channel gain between AP $j$ and user $i$. Also, $n_i[n]$$\sim$$\mathcal{CN}(0,\sigma^2_n)$ represents the received additive white Gaussian noise (AWGN) at user $i$, and $s_s[n]$ represents the sensing probing signal.

The $M\times1$ transmitted signals, $\mathbf{x}_j[n]$, at each AP \mbox{$j$$=$$1,...,N_{\text{Tx}}$} can be expressed as
\begin{align}\label{eq:transmitted-signal}
    \mathbf{x}_j[n] = \sum\limits_{i=1}^{N_{\text{UE}}}\mathbf{w}_{j,i}s_i[n] + \mathbf{w}_{j,t}s_s[n].
\end{align}

The received $M \times 1$ signal vector, $\mathbf{y}_r[n]$, at sensing receiver AP $r$ can be written as
\begin{equation}\label{eq:receiver-sensing}
    \mathbf{y}_r[n] = \sum\limits_{j=1}^{N_{\text{Tx}}}\alpha_{j,r}\sqrt{\beta_{j,r}}\mathbf{a}(\theta_r)\mathbf{a}(\theta_j)^T\mathbf{x}_j[n] + \mathbf{n}[n], 
\end{equation}

% swerling I källa
\noindent where $\mathbf{a}(\theta)$ is the $M\times 1$ antenna array response vector defined as $\mathbf{a}(\theta) =$$ \left[1 \quad e^{j\pi\cos{\theta}}\quad e^{j2\pi\cos{\theta}}\quad \dots \quad e^{j(M-1)\pi\cos{\theta}}\right]^T$ for a horizontal angle of arrival departure $\theta$. Further, in \eqref{eq:receiver-sensing} $\alpha_{j,r}\in\mathbb{C}$ is the radar cross section of the target following a Swerling-I model \cite{shnidman_expanded_2003} and $\beta_{j,r}$ is the channel gain between AP $j$, the target, and the sensing receiver AP. Also, $\mathbf{n}[n]\in\mathbb{C}^M$ is the AWGN vector with i.i.d. $\mathcal{CN}(0,\sigma_n^2)$ entries. The angles $\theta_j$ and $\theta_r$ are the angles of departure from transmitter AP $j$, and the reception angle of the echoes from the target to sensing receiver AP $r$, respectively. Note that the sample index, $n$, is omitted from here on out, and the following procedures are repeated for all $n$ and averaged, unless otherwise stated.
% Since we are considering a centralized cell-free MIMO ISAC-system with a joint control unit presumed aware of the transmit signal, interfering \textit{Line-of-Sight} (LOS) components and interference from \textit{Non-Line-of-Sight} (NLOS) paths caused by permanent stationary objects, may be calculated and cancelled at the receiver AP. 

\section{Adversary Model}
The adversary is assumed to be one of the communication users, which maliciously attempts to estimate the target’s position by inferring the directions of the main lobe in the beampatterns of the transmitting APs (their sensing beams). To this end, the adversary exploits the received sensing interference in an attempt to estimate the entire transmitted signal $\mathbf{x}_j$, and thereby, similarly to \cite{10587082,liu_joint_2020}, acquiring an estimate of the precoder, from the covariance matrix $\mathbf{R}_{\mathbf{x}_j}$$ =$$ \mathbf{x}_j \mathbf{x}_j^{\, H}$. By evaluating $\mathbf{a}(\theta_s)^H\mathbf{R}_{\mathbf{x}_j}\mathbf{a}(\theta_s)$ over an angular search area, $\theta_s$$\in$$ \Omega_s$, the adversary identifies the sensing beam’s direction as the angle that maximizes this expression. Since the locations of APs are known by the adversary, $\Omega_s$ is assumed to be the same quadrant as the correct angle, allowing the adversary to project a line from each AP location toward its corresponding $\theta_s$.

\subsection{Target Localization}
Herein, we propose a gradient descent method for estimating the closest point between the projected lines from each transmitting AP by solving the least squares problem that minimizes the loss function
\begin{equation}\label{eq:LS}
    \varepsilon (\mathbf{q}) = \dfrac{1}{N_l}\sum\limits_{i=1}^{N_l} ||\mathbf{q} - \mathbf{y}_i ||^2,
\end{equation}

\noindent where $\mathbf{q}$ is the estimated point, $N_l$ is the number of lines, and $\mathbf{y}_i$ is the closest point of each line to the estimated point. Then, if the target is within a predetermined radius of the estimated location, the adversary is deemed successful in determining the target position. 

For estimating the transmitted signal, the expectation-maximization algorithm \cite{moon-expectation-maximization-1996} is used. For AP $j$, the adversary considers the likelihood function $p(y_{j,a};\mathbf{x}_{j})$ where $y_{j,a}$ is the received signal at the adversary and $\mathbf{x}_{j}$ is the transmitted signal from the AP. Similarly to \cite{zhang-deep-2022}, we define $\mathbf{z}_{j,a}=\{\mathbf{h}_{j,a}, u\}$, as well as $\boldsymbol{\theta}=\{\mathbf{x}_{j}, \nu\}$, with $\nu$ being the degree of freedom of the underlying Student's $t$-distribution, and $u$ is an auxiliary latent variable. By establishing $q(\mathbf{z}_{j,a})$ as a variational arbitrary distribution, the expression to maximize becomes 
\begin{equation}
    \mathcal{F}(y_{j,a};q,\boldsymbol{\theta}) \!=\! \int q(\mathbf{z}_{j,a}) \log\left( \frac{p(y_{j,a}, \mathbf{z}_{j,a}; \boldsymbol{\theta})}{q(\mathbf{z}_{j,a})} \right) \, d\mathbf{z}_{j,a}.
\end{equation}

% As stated in \cite{zhang_deep_2022}, the likelihood function in~\eqref{eq:likelihood_function} can, after some assumptions and simplifications, be expressed as
% \begin{equation}\label{eq:double_integral_likelihood}
%     p(\widetilde{y}_{j,a};\bar{\mathbf{x}}_{j}) = \iint p(\mathbf{h}_{j,a}) \left(\mathbf{h}_{j,a} \bar{\mathbf{x}}_{j}, \frac{\sigma_a^2 \mathbf{I}}{u} \right) \mathcal{G}\left(\frac{\nu}{2}, \frac{\nu}{2} \right) \, d\mathbf{h}_{j,a} \, du.
% \end{equation}

% As a summary, the latent variables are now $\mathbf{h}_{j,a}=\{\mathbf{h}_{j,a}, u\}$, and the model parameters in the likelihood function are $\boldsymbol{\theta}=\{\bar{\mathbf{x}}_{j}, \nu\}$.

% To further simplify the likelihood function, a variational arbitrary distribution for the latent variable $\mathbf{h}_{j,a}$ called $q(\mathbf{h}_{j,a})$ is introduced, which in turn, from \cite{zhang_deep_2022}, allows one to define the optimizable parts as 
% \begin{equation}
%     \mathcal{F}(\widetilde{y}_{j,a};q,\boldsymbol{\theta}) = \int q(\mathbf{h}_{j,a}) \log\left( \frac{p(\widetilde{y}_{j,a}, \mathbf{h}_{j,a}; \boldsymbol{\theta})}{q(\mathbf{h}_{j,a})} \right) \, d\mathbf{h}_{j,a}.
% \end{equation}

\subsubsection{Expectation Step}
The expectation step maximizes $\mathcal{F}(y_{j,a};q,\boldsymbol{\theta})$, w.r.t. $q(\mathbf{z}_{j,a})$ which is achieved when
% \begin{equation}
%     \mathbb{E}_q[\mathbf{h}_{j,a}] = \left[\boldsymbol{\Omega}_{\mathbf{h}}\left(\frac{1}{\sigma_h^2} \widehat{\mathbf{h}}_{j,a}^H + \mathbf{x}_{j}\mathbf{x}_{j}^{\,H} \frac{\mathbb{E}_q[u]}{\sigma_n^2} \right) \right]^H,
% \end{equation}
\begin{align}
    &\mathbb{E}_q[\mathbf{h}_{j,a}] = \left[\boldsymbol{\Omega}_{\mathbf{h}}\left(\frac{1}{\sigma_h^2} \widehat{\mathbf{h}}_{j,a} + \mathbf{x}_{j}y_{j,a} \frac{\mathbb{E}_q[u]}{\sigma_n^2} \right) \right],\\
%\end{equation}
%\begin{equation}
    &\boldsymbol{\Omega}_{\mathbf{h}}= \left( \frac{1}{\sigma_h^2}\mathbf{I}_M + \mathbf{x}_{j}\mathbf{x}_{j}^{\,H} \frac{\mathbb{E}_q[u]}{\sigma^2_n} \right)^{-1},\\
%\end{equation}
%\begin{equation}\label{eq:u-mean}
    &\mathbb{E}_q[u] = \mathbb{E}\left[ \mathcal{G}\left( \frac{\nu+2}{2}, \frac{\nu+C}{2} \right) \right],\\
%\end{align}
%\begin{equation}
    &C = \frac{1}{\sigma^2_n}\left[\left(y_{j,a} - \mathbb{E}_q[\mathbf{h}_{j,a}]^H \mathbf{x}_{j} \right)^{\!2}\,  + M\mathbf{x}_{j}^H\boldsymbol{\Omega}_{\mathbf{h}} \mathbf{x}_{j}\right],
\end{align}

\noindent where $\mathcal{G}$ is the gamma distribution, and $\widehat{\mathbf{h}}_{j,a}$ and $\sigma_h^2$ denotes the initial channel estimate and its estimation error variance, respectively. For the dependency between $\mathbb{E}_q[u]$ and $\mathbb{E}_q[\mathbf{h}_{j,a}]$, we use the previous iteration's $\mathbb{E}_q[u]$ which is randomly initiated for the first iteration.

\subsubsection{Maximization Step}
The maximization step gives an estimate of $\mathbf{x}_{j}$ by maximizing $\mathcal{F}(y_{j,a};q,\boldsymbol{\theta})$ with regards to $\boldsymbol{\theta}$. For transmitter AP $j$ the estimate is given by
\begin{equation}\label{eq:maximization_step}
    \widehat{\mathbf{x}}_{j} = \argmin_{\mathbf{x}_j} \left[ (y_{j,a} - \mathbb{E}_q[\mathbf{h}_{j,a}]^H \mathbf{x}_j)^2 + M\mathbf{x}_j^{\,H} \boldsymbol{\Omega}_{\mathbf{h}} \mathbf{x}_j \right],
\end{equation}

\noindent and has a closed form solution detailed in \cite{zhang-deep-2022}.

\section{Privacy-Aware Framework}
We propose a system design based on reducing the sensing signal leakage by minimizing the sum MI of the sensing signal and the received signal at the users. From \cite{marzetta-fundamentals-2016}, the MI can be upper bounded by 
\begin{equation}\label{eq:MI}
    I(y_i;\mathbf{x}^s_j) \leq \log_2 \left( 1+\left|\mathbf{h}^H_{j,i}\mathbf{x}^s_j\right|^2\right),
\end{equation}
\noindent where $\mathbf{h}_{j,i}\in\mathbb{C}^{M}$ is the channel vector between AP $j$ and user $i$, and $\mathbf{x}_j^s=\mathbf{w}_{j,t}s_s$. The sum MI is minimized by
\begin{equation}\label{eq:MI-simple}
    \min_{\mathbf{W}} \quad \sum\limits_{i=1}^{N_{\text{UE}}} \sum\limits_{j=1}^{N_{\text{Tx}}} \sum\limits_{n=1}^{N}\left|\mathbf{h}^H_{j,i}\mathbf{x}^s_j\right|^2\\
\end{equation}

\noindent where $\mathbf{W}$ is a $M N_\text{Tx}\times (N_\text{UE}+1)$ matrix consisting of each transmitter APs $M \times (N_\text{UE}+1)$ precoder matrix, $\mathbf{W}_j$ for AP $j=1,2,...,N_\text{Tx}$. Hence, the aggregate precoder matrix \mbox{$\mathbf{W} = \left[ \mathbf{W}_1^T, \mathbf{W}_2^T,...,\mathbf{W}_{N_\text{Tx}}^T \right]^T$} columns represent the total precoding vector for a user from every transmitter AP, where the last column represents the target.

\subsection{Precoder Design}
Given a set of APs with a pre-assigned role, either transmitter or sensing receiver, we formulate the following optimization problem to find the transmit precoder.
\begin{subequations}\label{eq:precoder-op}
    \begin{alignat}{2}
    \min_{\mathbf{W}} \quad & \sum\limits_{i=1}^{N_{\text{UE}}} \sum\limits_{j=1}^{N_{\text{Tx}}} \sum\limits_{n=1}^{N}\left|\mathbf{h}^H_{j,i}\mathbf{x}^s_j\right| ^2\\
    \textrm{s.t.} \quad & \gamma_s \geq \gamma_1\\
    & \gamma_{\text{UE}_i} \geq \gamma_2, \forall i \label{eq:ue-sinr-constraint}\\
    & P_{j} \leq P_{max}, \forall j, \label{eq:pow-constraint}
    \end{alignat}
\end{subequations}

 \noindent wherein, $\gamma_s$ is the sensing SINR, and $\gamma_1$ is a minimum sensing SINR constraint. Further, $\gamma_{\text{UE}_i}$ and $\gamma_2$ represents the received SINR per user and the minimum SINR threshold for a user, respectively. $P_{j}$ is the transmit power from AP $j$, and $P_{max}$ is the maximum allowed transmission power for each antenna.

From \eqref{eq:receiver-sensing}, we can derive the SINR expression of the sensing signal as 
\begin{align}\label{eq:objective-function}
 \nonumber  & \gamma_s \!= \!\frac{\displaystyle \!\!\!\sum\limits_{r=1}^{N_{\text{Rx}}} \sum\limits_{n=1}^N \!\mathbf{s}^H 
        \mathbf{W}^H \!\mathbf{A}_r \!\mathbf{W}
    \mathbf{s}
    }
    {\displaystyle N_{\text{Rx}}NM\sigma^2_n},
%\end{equation}
%\[
\,\, \text{with}\\ \,\,
&\mathbf{A}_r \!=\!\! \!\begin{bmatrix} 
    \mathbf{A}_{1,1} & \!\!\!\!\dots\!\!  & \mathbf{A}_{1,N_\text{Tx}}\\
    \vdots &\!\!\!\! \!\!\ddots\!\! & \vdots\\
    \mathbf{A}_{N_\text{Tx}, 1} &\!\! \!\dots\!  & \mathbf{A}_{N_\text{Tx},N_\text{Tx}} 
    \end{bmatrix} \!\!\in\!\mathbb{C}^{M N_\text{Tx} \times M N_\text{Tx}},
%\]
\end{align}

\noindent with $\mathbf{A}_{j,m}$ being the $M \times M$ matrices of transmitter AP $j$ and $m$ defined by  
\begin{equation*}
    \mathbf{A}_{j,m} = \sqrt{\beta_{j,r}\beta_{m,r}}\mathbf{a}(\theta_j)^*\mathbf{a}(\theta_r)^H  \alpha_{j,r}^* \alpha_{m,r} \mathbf{a}(\theta_r) \mathbf{a}(\theta_m)^T.
\end{equation*}

\noindent In \eqref{eq:objective-function}, $\mathbf{s} = \left[ s_1[n],...,s_{N_\text{UE}}[n],s_s[n]\right]^T$. To formulate \eqref{eq:precoder-op} as a convex optimization problem, a first order Taylor approximation is applied to the term $\mathbf{W}^H \mathbf{A}_r \mathbf{W}$ around the result of the previous iteration of the procedure, resulting in
\begin{equation}\label{eq:taylor-obj-fun}
    P_1^r(\mathbf{W}_p) = 2\mathbf{W}^H_{p-1}\mathbf{A}_r\left( \mathbf{W}_p - \mathbf{W}_{p-1} \right) + \mathbf{W}^H_{p-1} \mathbf{A}_r \mathbf{W}_{p-1}.
\end{equation}

\noindent and thus a final expression for the objective function is obtained as
\begin{equation}\label{eq:s-sinr-final}
    \gamma_s = \frac{\displaystyle  \sum\limits_{r=1}^{N_{\text{Rx}}} \sum\limits_{n=1}^N \mathbf{s}^H 
        P_1^r(\mathbf{W})
    \mathbf{s}
    }
    {\displaystyle N_{\text{Rx}}NM\sigma^2_n}.
\end{equation}

The communication SINR constraint is reformulated as
\begin{equation}\label{eq:ue-sinr-final}
    \gamma_{\text{UE}_i} =  \frac{\displaystyle  
            \left| \mathbf{h}_i^H \mathbf{w}_i \right| ^2
    }
    {\displaystyle 
        \biggl| \sum\limits_{\substack{k=1 \\ k\ne i}}^{N_\text{UE}} \mathbf{h}_{i}^H \mathbf{w}_{k} \biggr| ^2 + 
        \left| \mathbf{h}_{i}^H \mathbf{w}_{t} \right| ^2 + 
        \sigma^2_n
    },
\end{equation}

\noindent by using the substitutions $\mathbf{h}_i = \left[ \mathbf{h}^T_{1,i}, \mathbf{h}^T_{2,i},...,\mathbf{h}^T_{N_\text{Tx},i} \right]^T \in \mathbb{C}^{MN_\text{Tx}}$ and $\mathbf{w}_i = \left[ \mathbf{w}^T_{1,i}, \mathbf{w}^T_{2,i},...,\mathbf{w}^T_{N_\text{Tx},i} \right]^T \in \mathbb{C}^{MN_\text{Tx}}$, representing the concatenated channel vector and precoder vector to user $i$ (similarly for users $k$, and target $t$) from all the transmitting APs, respectively.  Again, a first order Taylor approximation is applied around the previous iteration's value of $\mathbf{w}_i$. Also, the slack variables $\tau_d$ and $\tau_n$ for the denominator and numerator in \eqref{eq:ue-sinr-final}, respectively, are introduced. The final expression for the constraint in \eqref{eq:ue-sinr-constraint} is therefore
\begin{subequations}\label{eq:ue-sinr-constraint-final}
    \begin{multline}
        (\mathbf{w}_i^H)_{(p-1)}\mathbf{h}_i\mathbf{h}_i^H(\mathbf{w}_i)_{(p-1)} \\
        + 2(\mathbf{w}_i^H)_{(p-1)}\mathbf{h}_i\mathbf{h}_i^H((\mathbf{w}_i)_p - (\mathbf{w}_i)_{(p-1)}) \geq \tau_n
    \end{multline}
\vspace{-0.3cm}
    \begin{equation}
        \biggl| \sum\limits_{\substack{k=1 \\ k\ne i}}^{N_\text{UE}} \mathbf{h}_{i}^H \mathbf{w}_{k} \biggr| ^2 + 
        \left| \mathbf{h}_{i}^H \mathbf{w}_{t} \right| ^2 + 
        \sigma^2_n \leq \tau_d
    \end{equation}
    
    \begin{equation}
        \tau_n \geq \tau_d \gamma_2.
    \end{equation}    
\end{subequations}

\subsection{Access Point Selection}
The second step of the proposed method to counteract the adversarial user, involves a minimization of the MI w.r.t. APs being ISAC-signal receivers or transmitters. Consider the set of all APs, $\mathcal{A}$$=$$\{a_l\}_1^{N_{\text{AP}}}$, where $a_l$ is each AP $l=1,2,...,N_{AP}$, and $N_{AP}$ is the total number of APs, both receiving and transmitting. Thus, the problem in~\eqref{eq:MI-simple} is formulated as
\begin{equation}\label{eq:selection-opt}
    \min_{\mathcal{R}\subset \mathcal{A}} \quad \sum\limits_{i=1}^{N_{\text{UE}}} \sum\limits_{l=1}^{N_{\text{AP}}}\sum\limits_{n=1}^{N} \left|\mathbf{h}_{l,i}^H\mathbf{x}^s_l\right| ^2,
\end{equation}

\noindent where $\mathcal{R}$ is the set of receiver APs, and $\mathbf{x}^s_l\in\mathbb{C}^{M}$ is the precoded sensing signal from each AP $l$, i.e. $\mathbf{x}^s_l = \mathbf{w}_{l,t}s_s$.

By defining $\mathbf{l}_{\text{Rx}} = \left[l_1,l_2,...,l_{N_{\text{Rx}}}\right]^T$ as the unique indexes of the receiver APs in $\mathcal{A}$, and the matrix
\[
\mathbf{M} = \sum\limits_{n=1}^{N}\begin{bmatrix} 
    \left|\mathbf{h}^H_{1,1}\mathbf{x}^s_1\right| ^2 & \dots  & \left|\mathbf{h}^H_{N_{\text{AP}},1}\mathbf{x}^s_{N_{\text{AP}}}\right|^2 \\
    \vdots & \ddots & \vdots\\
     \left|\mathbf{h}^H_{1,N_{\text{UE}}}\mathbf{x}^s_1\right|^2 & \dots  & \left|\mathbf{h}^H_{N_{\text{AP}},N_{\text{UE}}}\mathbf{x}^s_{N_{\text{AP}}}\right|^2 
    \end{bmatrix},
\]

\noindent the solution to \eqref{eq:selection-opt} becomes
\begin{equation}\label{eq:selection-op-final}
    \mathcal{R} = \left\{a_l\in \mathcal{A} \,\, \forall l\in\mathbf{l}_{\text{Rx}} : \argmax_{\mathbf{l}_{\text{Rx}}} \sum\limits_{l\in\mathbf{l}_{\text{Rx}}}\left|\left|\mathbf{m}_l\right|\right|\right\},
\end{equation}
\noindent with $\mathbf{m}_l$ being the $l$th column vector of $\mathbf{M}$.

\section{Results}
The results presented in this section are produced using the parameters and code in\footnote{\href{https://github.com/diamo98/ISAC-group-LiU/tree/main/Privacy-preservation\%20in\%20ISAC/Privacy-Preserving\%20Framework\%20for\%20Cell-Free\%20MIMO\%20ISAC\%20Systems/privacy\_for\_isac}{https://github.com/ISAC-group-LiU/Privacy-for-ISAC}}. The simulated system consists of four APs, whereas one of them is a sensing receiver. The users, including the adversary, are placed within an area enclosed by the APs, with the target being directly in the middle of the grid area.
% \footnote{\href{https://github.com/diamo98/ISAC-group-LiU/tree/main/Privacy-preservation\%20in\%20ISAC/Privacy-Preserving\%20Framework\%20for\%20Cell-Free\%20MIMO\%20ISAC\%20Systems/privacy\_for\_isac}{https://github.com/ISAC-group-LiU/Privacy-preservation}}
% \begin{table}[bt]
% \caption{Simulation parameters.}
%     \centering
%     \resizebox{\linewidth}{!}{%
%     \begin{tabular}{| c | c || c | c |}
%         \hline
%         \textbf{Parameter} & \textbf{Value} & \textbf{Parameter} & \textbf{Value} \\\hline
%         Grid Area & $1000 \times 1000$ & $N_{\text{UE}}$ & $3$\\\hline
%         Modulation Order & $16$ & $N_{\text{AP}}$ & $4$\\\hline
%         Path Loss Exponent & $3$ & $N$ & $16$\\\hline 
%         $R_{\text{GD}}$ &  $10$ & $P_{max}$ & $35$ dBm\\\hline
%         $\sigma_n$ &  $-94$ dBm & $\gamma$ & $3$ dB\\\hline
%         $\eta$ & $0.75$ & $\sigma_H$ & $-40$ dB\\\hline
%         $M$ & $64$ & $\alpha_{j,r}$ & $10$ dB\\\hline
%     \end{tabular}%
%     }
%     \label{tab:parameter-table}
% \end{table}

%\begin{figure}[bt]

\begin{figure}
\centering
\usetikzlibrary{plotmarks}

\definecolor{mycolor1}{rgb}{0.0, 0.45, 0.70}%{0.090, 0.250, 0.500}
\definecolor{mycolor2}{rgb}{0.90, 0.55, 0.15}%{0.950, 0.600, 0.100}
\begin{tikzpicture}
\begin{axis}[%
      grid=both, % draw both major and minor grids
      major grid style={gray!40, thin}, % faint gray major grid
      minor grid style={gray!10, thin}, % even fainter minor grid
      minor tick num = 1,
      xmin=0.5,xmax=15,
      ymin=0,ymax=25,
      axis y line*=left,
      xlabel={$\dfrac{\gamma_1}{\gamma_2}$},
      xlabel near ticks,
      ylabel={\color{mycolor2}Sensing SINR, $\gamma_s$},
      ylabel near ticks,
      y tick style={color=mycolor2},
      y axis line style={color=mycolor2},
      y tick label style={color=mycolor2},
      legend style={
        at={(0.5,1.1)},
        anchor=north,
        legend columns=2,      % put legend entries side by side
        /tikz/every even column/.append style={column sep=0.5cm} % extra spacing if desired
      },
      legend image post style={xscale=1}, % make legend lines longer
      legend cell align={left}
    ]
\addplot [color=mycolor2, densely dotted, line width=1.5pt, forget plot]
  table[row sep=crcr]{%
0.5	21.536128664641\\
1	21.536128664641\\
2	21.536128664641\\
4	21.536128664641\\
10	21.536128664641\\
15	21.536128664641\\
20	21.536128664641\\
};
\addlegendimage{thick, densely dotted, black}
\addlegendentry{max $\gamma_s$}

\addplot [color=mycolor2, loosely dashed, line width=1.5pt, forget plot]
  table[row sep=crcr]{%
0.5	21.1681864137076\\
1	21.1681864137076\\
2	21.1681864137076\\
4	21.1681864137076\\
10	21.1681864137076\\
15	21.1681864137076\\
20	21.1681864137076\\
};
\addlegendimage{thick, loosely dashed, black}
\addlegendentry{max $\gamma_s$ with selection}

\addplot [color=mycolor2, line width=1.5pt, mark size=3.0pt, mark=triangle, mark options={mycolor2}, forget plot]
  table[row sep=crcr]{%
0.5	2.81747565575782\\
1	6.24839374238641\\
2	10.0259846185833\\
4	14.4141413266716\\
10	18.7323712949196\\
15	20.591314014585\\
20	22.0012116773423\\
};
\addlegendimage{thick, black, mark=triangle}
\addlegendentry{min MI}

\addplot [color=mycolor2, line width=1.5pt, mark size=3.0pt, mark=x, mark options={mycolor2}, forget plot]
  table[row sep=crcr]{%
0.5 2.78259325380480\\
1 6.38318682104090\\
2 10.4892265796628\\
4 14.2959065759084\\
10 18.7455298326967\\
15 20.6022412133346\\
20 21.6944170754655\\
};
\addlegendimage{thick, black, mark=x}
\addlegendentry{min MI with selection}

\end{axis}

\begin{axis}[%
      minor tick num = 1,
      xmin = 0.5, xmax = 20,
      ymin = 0, ymax = 1,
      hide x axis,
      axis y line*=right,
      ylabel={\color{mycolor1}Probability of Detection, $P_D$},
      ylabel near ticks,
      y tick style={color=mycolor1},
      y axis line style={color=mycolor1},
      y tick label style={color=mycolor1}    
      ]
\addplot [color=mycolor1, densely dotted, line width=1.5pt]
  table[row sep=crcr]{%
0.5	0.489333333333333\\
1	0.489333333333333\\
2	0.489333333333333\\
4	0.489333333333333\\
10	0.489333333333333\\
15	0.489333333333333\\
20	0.489333333333333\\
};

\addplot [color=mycolor1, loosely dashed, line width=1.5pt]
  table[row sep=crcr]{%
0.5	0.325333333333333\\
1	0.325333333333333\\
2	0.325333333333333\\
4	0.325333333333333\\
10	0.325333333333333\\
15	0.325333333333333\\
20	0.325333333333333\\
};

\addplot [color=mycolor1, line width=1.5pt, mark size=3.0pt, mark=triangle, mark options={mycolor1}]
  table[row sep=crcr]{%
0.5	0.00181159420289855\\
1	0.0108695652173913\\
2	0.0307971014492754\\
4	0.0869565217391304\\
10	0.152173913043478\\
15	0.206521739130435\\
20	0.230072463768116\\
};

\addplot [color=mycolor1, line width=1.5pt, mark size=3.0pt, mark=x, mark options={mycolor1}]
  table[row sep=crcr]{%
0.5 0\\
1 0.00429184549356223\\
2 0.0214592274678112\\
4 0.0472103004291846\\
10 0.124463519313305\\
15 0.152360515021459\\
20 0.150214592274678\\
};

\end{axis}
\end{tikzpicture}%
\caption{Numerical results of the proposed framework for a fixed system layout. Data points were generated by averaging 1000 trials.}\label{fig:results-fig}
\end{figure}

Figure \ref{fig:results-fig} shows the sensing SINR, and the probability of the target being detected by the adversary, as a function of the fraction ${\gamma_1}/{\gamma_2}$, which determines how much power is spent on the sensing task relative to the power for communication. Experiments were carried out over several trials and $P_D$ corresponds to the amount of trials where the adversary successfully determined the target location. A variant of the precoder where $\gamma_s$ was maximized, as opposed to MI minimized, serves as an optimal sensing performance comparison. \textit{Selection} refers to the AP selection described as the second block of the proposed framework. In Figure \ref{fig:results-fig}, the effects of different individual components of the framework are visualized, and the gain from each block is clear, i.e., improvement in $P_D$ is observed from first only applying the proposed precoder, but also from the combination of precoder and selection. There is a tradeoff between user service and privacy observed in \ref{fig:results-fig}, where a low ratio $\frac{\gamma_1}{\gamma_2}$ yields and improved privacy gain, but still at maximum sensing SINR, the framework yields a privacy gain. As reduction in $P_D$ gradually lessens closer to 0\%, distinguishing the effectiveness of each individual contribution of the two MI-preserving blocks is difficult. A more refined threat model, capable of a higher probability of target detection than 50\% in the ideal case, could provide further insights of the effectiveness of the framework. It is also important to note that while the framework has the same SINR performance with and without selection, the computational load increases significantly with the inclusion of the selection procedure.

\section{Conclusions}
We propose a novel two-block, precoder and AP operation selection, framework seeking to reduce precoder information inadvertently present in ISAC signals. For this purpose, the MI of the transmitted sensing signal, and the signal received by each user, is minimized. Using an attack model where the leaked precoder information is exploited to locate a radar detection target, we show that the framework reaches the same level of sensing performance as a privacy-oblivious system, while yielding a reduction in the success rate of the proposed attack.

\vfill\pagebreak

\section{Funding Acknowledgments}
This publication is based upon work from COST Action 6G-PHYSEC (CA22168), supported by COST (European Cooperation in Science and Technology). The work has also been partially supported by the Swedish strategic research environment ELLIIT and, as well supported by Portuguese FCT – Fundação para a Ciência e a Tecnologia, I.P., under project reference UID/50008/2023 IT.

\bibliographystyle{IEEEbib}
\bibliography{references}

\end{document}